\documentclass[12pt]{elsarticle}

\usepackage[T1]{fontenc}

\usepackage{amssymb}
\usepackage{amsmath}

\usepackage[breaklinks=true, pdftex, 
pdfborder={0 0 0}, colorlinks=true, linkcolor=blue, citecolor=blue, urlcolor=blue, 
bookmarks=false, pdfpagelabels=false]{hyperref}
\usepackage{comment}
\usepackage{xcolor}
\usepackage{multirow,makecell}

\newcommand{\review}[1]{{{#1}}}

\usepackage{xspace}

\newcommand{\R}{\mathbb{R}}
\renewcommand{\vec}[1]{\mathbf{#1}}
\newcommand{\evec}{\vec{e}}
\newcommand{\xvec}{\vec{x}}
\newcommand{\Jvec}{\vec{J}}
\newcommand{\rvec}{\vec{r}}

\newcommand{\np}{n_x}
\newcommand{\nd}{d}
\newcommand{\nn}{n_n}
\newcommand{\dataset}{\mathcal{D}}
\newcommand{\xHat}{\hat{\xvec}}

\newcommand{\rhonm}{$\rho_c$\xspace}
\newcommand{\rhonmcode}{\code{RHO_NM}\xspace}
\newcommand{\eovera}{$E^{\mathrm{NM}}/A$\xspace}
\newcommand{\eoveracode}{\code{E_NM}\xspace}

\newcommand{\knmcode}{\code{K_NM}\xspace}
\newcommand{\ass}{$a^{\mathrm{NM}}_{\mathrm{sym}}$\xspace}
\newcommand{\asscode}{\code{ASS_NM}\xspace}
\newcommand{\lass}{$L^{\mathrm{NM}}_{\mathrm{sym}}$\xspace}
\newcommand{\lasscode}{\code{LASS_NM}\xspace}
\newcommand{\smasscode}{\code{SMASS_NM}\xspace}
\newcommand{\crdrz}{$C^{\rho\Delta\rho}_0$\xspace}
\newcommand{\crdro}{$C^{\rho\Delta\rho}_1$\xspace}
\newcommand{\crdrcode}{\code{CrDr}\xspace}
\newcommand{\cpvzz}{$V^n_0$\xspace}
\newcommand{\cpvzo}{$V^p_0$\xspace}
\newcommand{\cpvzcode}{\code{CpV0}\xspace}
\newcommand{\crdjz}{$C^{\rho\nabla J}_0$\xspace}
\newcommand{\crdjo}{$C^{\rho\nabla J}_1$\xspace}
\newcommand{\crdjcode}{\code{CrDJ}\xspace}
\newcommand{\cjcode}{\code{CJ}\xspace}

\newcommand{\hfbtho}{HFBTHO\xspace}
\newcommand{\hfbthoad}{HFBTHO-AD\xspace}
\newcommand{\slyf}{SLy4\xspace}
\newcommand{\unedfnb}{UNEDFnb\xspace}
\newcommand{\unedfpre}{UNEDFpre\xspace}
\newcommand{\ecnoise}{ECNoise\xspace}

\usepackage{listings}
\newcommand*{\code}{\lstinline[basicstyle=\ttfamily,breaklines=true, breakatwhitespace=false,prebreak=-]}
\makeatletter
\def\lst@lettertrue{\let\lst@ifletter\iffalse}
\makeatother
\usepackage{adjustbox}

\usepackage{marginnote}
\usepackage{etoolbox}
\makeatletter
\patchcmd{\@mn@@@marginnote}{\begingroup}{\begingroup\@twosidefalse}{}{\fail}
\makeatother
\usepackage{cleveref}
\usepackage{orcidlink}

\date{}
\journal{Computer Physics Communications}

\begin{document}

\begin{frontmatter}



\title{\hfbthoad: Differentiation of a nuclear energy density functional code}


\author[inria]{Laurent Hasco\"et\orcidlink{0000-0002-5361-0713}}
\author[argonne]{Matt Menickelly\orcidlink{0000-0002-2023-0837}}
\author[argonne]{Sri Hari Krishna Narayanan\orcidlink{0000-0003-0388-5943}\corref{cor1}} 
\author[argonne]{Jared O'Neal\orcidlink{0000-0003-2603-7314}}
\author[livermore]{Nicolas Schunck\orcidlink{0000-0002-9203-6849}}
\author[berkeley]{Stefan M. Wild\orcidlink{0000-0002-6099-2772}}

\cortext[cor1]{Corresponding Author}
\affiliation[inria]{organization={INRIA, Sophia-Antipolis},
            addressline={2004 Route des lucioles}, 
            city={Valbonne},
            postcode={BP 93 06902}, 
            country={France}}
            
\affiliation[argonne]{organization={Argonne National Laboratory},
            addressline={9700 S. Cass Ave.}, 
            city={Lemont},
            state={IL},
            postcode={60439}, 
            country={USA}}

\affiliation[livermore]{organization={Lawrence Livermore National Laboratory},
            addressline={7000 East Ave.}, 
            city={Livermore},
            state={CA},
            postcode={94550}, 
            country={USA}}
            
\affiliation[berkeley]{organization={Lawrence Berkeley National Laboratory},
            addressline={1 Cyclotron Road}, 
            city={Berkeley},
            state={CA},
            postcode={94720},
            country={USA}}

\begin{abstract}
The \hfbtho code implements a nuclear energy density functional solver to model the structure of atomic nuclei. \hfbtho has previously been used to calibrate energy functionals and perform sensitivity analysis by using derivative-free methods. To enable derivative-based optimization and uncertainty quantification approaches, we must compute the derivatives of \hfbtho outputs with respect to the parameters of the energy functional, which are a subset of all input parameters of the code. We use the algorithmic/automatic differentiation (AD) tool Tapenade to differentiate \hfbtho. We compare the derivatives obtained using AD against finite-difference approximation and examine the performance of the derivative computation.
\end{abstract}


\end{frontmatter}

\section{Introduction}
\label{sec:introduction}

Nuclear energy density functional (EDF) methods aim to provide complete characterization of properties of atomic nuclei \cite{schunck2019energy}. Similar in spirit to the density functional theory (DFT) of electrons in atoms and molecules \cite{parr1989density,dreizler1990density,eschrig1996fundamentals}, it is a computationally efficient reformulation of the quantum many-body problem of $A$ nucleons (protons and neutrons) interacting through in-medium effective potentials such as the Skyrme \cite{stone2007skyrme} or Gogny \cite{robledo2019mean} force or their relativistic equivalent \cite{niksic2011relativistic}. Computational implementations of EDF methods are based on either  basis expansion methods or lattice techniques. Examples of the former include the HOSPHE \cite{carlsson2010solution}, \hfbtho \cite{stoitsov2005axially,stoitsov2013axially,perez2017axially,marevic2022axiallydeformed}, HFODD \cite{dobaczewski1997solution,dobaczewski1997solutiona,dobaczewski2000solution,dobaczewski2004solution,dobaczewski2009solution,schunck2012solution,schunck2017solution,dobaczewski2021solution}, and DIRHB \cite{niksic2014dirhb} solvers, which expand solutions on the eigenfunctions of the quantum harmonic oscillator; examples of the latter include HFBRAD \cite{bennaceur2005coordinatespace}, EV8 \cite{bonche2005solution,ryssens2015solution}, SkyAX \cite{reinhard2021axial}, and solvers that discretize space within a box \cite{pei2014adaptive,chen2022threedimensional}.

Nuclear EDF theory has a few significant differences from electronic DFT \review{\cite{parr1989density,dreizler1990density,engel2007intrinsicdensity,messud2009density}}. In contrast to the potentials for systems of electrons in atoms or molecules---the Coulomb force, which is of a two-body nature only---the nuclear potential contains in principle two-body, three-body, and up to $A$-body terms. However, this expansion is not  well known. In practice, most nuclear energy density functionals are constructed from empirical models of two-body effective potentials (e.g., the Skyrme and Gogny potentials). In addition, the small number of constituents---protons and neutrons---in atomic nuclei means that 
\review{beyond mean-field} correlations can be \review{significant}. The formal consequence is that nuclear EDF methods rely on spontaneous symmetry breaking to effectively incorporate such correlations into the simple EDF scheme \review{---a technique that has also been recently leveraged by ab initio approaches \cite{duguet2015symmetry,duguet2016symmetry,yao2018generatorcoordinate,hagen2022angularmomentum,hu2024abinitio,sun2025multiscale}}. The practical consequence of symmetry breaking is that the nuclear wavefunction will not obey symmetries of the nuclear Hamiltonian (and will not conserve the associated quantum numbers): nuclei can be deformed, their number of particles may not be constant, and so on. Most of the available EDF solvers differ by which symmetries are built into the code by default. In this work we focus on the \hfbtho solver, which assumes axial and time-reversal symmetry but allows for the breaking of reflection symmetry and particle number.

Because nuclear energy functionals are phenomenological constructs, they depend on a small (on the order of 5--20 depending on the model) number of parameters that must be calibrated on experimental data. \review{These experimental data typically include binding energies of spherical or well-deformed nuclei, proton radii, energy differences between even-even and odd nuclei, the excitation energy of fission isomers or the energy of the giant dipole resonance, etc.; see Chapter 9 of \cite{schunck2019energy} for a discussion. Until about 15 years ago, the objective function encoding the discrepancy between experimental observables and theoretical predictions often included data in simple systems such as nuclear matter or spherical, closed-shell nuclei because of computational limitations. The Universal Nuclear Energy Density Functional (UNEDF) collaboration \cite{bogner2013computational} was the first to perform a fully-fledged optimization with an objective function including dozens of deformed nuclei treated the the Hartree-Fock-Bogoliubov approximation \cite{kortelainen2010nuclear,kortelainen2012nuclear,kortelainen2014nuclear,wild2015derivativefree}.}

Because the outputs of the EDF did not include derivatives of the various observables with respect to the calibration's decision parameters, this prior optimization work used derivative-free optimization \cite{LMW2019AN}. A potential improvement would be to use derivative-based optimization methods~\cite{nocedal2006no}; however, using such methods would require derivatives of the outputs with respect to the parameters. Such derivatives could also provide valuable for other higher-level tasks such as sensitivity analysis and uncertainty quantification (UQ) \cite{Smith2014,carlsson2016uncertainty}.

To formulate the derivatives that must be computed, we consider the $\nd (= 108)$ experimental values associated with $\nn (= 72)$ different nucleus configurations in the UNEDF0
dataset~\cite{kortelainen2010nuclear},
\[
\dataset = \{y_i\}_{i=1}^{\nd},
\]
and the weighted least-squares objective function used for the UNEDF0 optimization, 
\begin{equation}
f(\xvec) = ||\rvec(\xvec)||_2^2 = \sum_{i=1}^{\nd} r_i^2.
\label{eq:ObjectiveFunction}
\end{equation}
Here $\rvec = (r_1, \cdots, r_{\nd}) : \R^{\np} \to \R^{\nd}$ is the weighted residual function \review{for $\np (= 12)$ input parameters}, where the individual residuals evaluated at a given parameter point $\xvec \in \R^{\np}$ are defined as
\[
r_i(\xvec) = \frac{y_i - m_i(\xvec)}{w_i}.
\]
The $m_i : \R^{\np} \to \R$ are the model functions, and the $w_i$ are positive weights assigned to each residual; an example setting of the $w_i$ appropriate for the UNEDF0 dataset is provided in~\cite{UNEDF0}. The $d$ many $m_i(\xvec)$ values are computed by executing \hfbtho once for each of the $\nn$ nucleus configurations. 
Each computed $r_i(\xvec)$ quantifies the normalized deviation between the theoretical value, $m_i(\xvec)$, 
and the observable value, $y_i$, in $\dataset$. Thus $f(\xvec)$ 
provides a scalar value that expresses how well the theory fits the data at the parameter point $\xvec$. 

To enable derivative-based optimization, we construct the Jacobian function $\Jvec$ that returns the Jacobian matrix $\Jvec(\xvec)\in \R^{ \nd \times \np}$ of $\rvec$ at $\xvec$.  Based on the least-squares structure of the objective function $f$, we can also determine immediately the gradient of the objective function at $\xvec$ using
\begin{equation*}
\nabla f(\xvec) = 2 \Jvec(\xvec)^T\rvec(\xvec).
\label{eq:grad_f}
\end{equation*}

Numerical differentiation methods, such as finite differencing,  estimate derivatives using just the original function code. For example, to numerically approximate the partial derivatives of $\Jvec$, we can use
forward differences to compute the elements of $\Jvec$ via
\begin{equation}
J_{ij}(\xvec)=\frac{\partial r_i}{\partial x_j}(\xvec) \approx \frac{r_i(\xvec+h_{ij}\evec_j)-r_i(\xvec)}{h_{ij}},
\label{eqn:fd}
\end{equation}
where $\evec_j$ is the unit vector of all zeros except for a one in the $j$th
entry and $h_{ij}>0$ is the difference parameter.  The decision to use only a first-order finite-difference (FD) scheme is
motivated by the desire to use the fewest number of residual function
evaluations possible. 

In addition to the truncation error (associated with nonzero higher-order derivatives), the quality of these approximate derivatives can suffer from round-off noise in the EDF computation.
The composition of round-off with finitely terminated iterative procedures (e.g., for eigenvalue approximation) within a residual function calculation results in deterministic computational noise, which can make obtaining reliable finite-difference approximations difficult in practice.
If $h_{ij}$ is chosen too small, then the signal of the difference is lost to the noise. 
On the other hand, if $h_{ij}$ is chosen too large, then the finite-difference estimate fails to approximate the derivative well (due to the nonzero higher-order derivatives). 
Techniques for choosing $h_{ij}$ to achieve a balance between noise and truncation error have been established (e.g., \ecnoise \cite{more2011edn}) but typically require multiple evaluations of $\rvec$ at different points $\xvec$.

For large complex codes such as \hfbtho, analytical computation of derivatives can be a time-consuming and error-prone task. Moreover, as the code changes, the code that computes the derivatives must also be updated.

An alternative to approximation methods or laborious hand coding is to utilize algorithmic/automatic differentiation (AD) to compute the derivatives~\cite{Griewank2008EDP,NaumannBook,NARAYANAN20101845}. 
\review{AD computes the derivatives of the outputs of a mathematical function, expressed as source code, with respect to any independent parameters, provided that a differentiable relationship exists between these parameters and the outputs.}
AD can be implemented at compile time or at run time. 
AD relies on the larger computation being broken down into intrinsic mathematical operations. By differentiating each operation and combining their \review{partial} derivatives using the chain rule of calculus, the derivatives for the large computation can be computed. If the \review{partial} derivatives are combined from the first operation to the last, it gives rise to the so-called \emph{forward mode} of AD. When the derivatives are combined from the last to the first, it gives rise to the so-called \emph{reverse mode} of AD.
\review{For a function $g : \R^{N} \to \R^{M}$, the forward mode computes the derivatives at the cost of $\mathrm{O}(N)$ function evaluations due to the direction of combination of partial derivatives of the operations. The reverse mode similarly computes the derivatives at the cost of $\mathrm{O}(M)$ function evaluations. This implies that the forward mode
is more efficient when $N<M$ and the reverse mode is more efficient otherwise.}
The reverse mode, also called backpropagation, \review{therefore}, underpins supervised machine learning techniques \cite{2015arXiv150205767G}. 
AD can use compiler-based techniques to generate a new source code in languages such as Fortran and C~\cite{Hascoet2013,Giering1998} or can use techniques such as operator overloading in languages such as C++~\cite{10.1145/229473.229474,SaAlGauTOMS2019}. AD has been implemented at the intermediate representation level inside the LLVM compiler by the tool Enzyme~\cite{NEURIPS2020_9332c513}. AD has also been implemented in Python-based machine learning frameworks such as JAX~\cite{jax2018github}, PyTorch~\cite{10.5555/3454287.3455008}, and TensorFlow~\cite{tensorflow2015-whitepaper} as well as  for Julia~\cite{10.1145/3458817.3476165}.

The main contribution of this work is to show how to efficiently compute the derivatives of \hfbtho using AD. We use the open-source AD engine Tapenade \cite{Hascoet2013} to differentiate the \hfbtho code that computes the model functions $m_i$, thus generating the $m_i'$ that are then composed using the chain rule of calculus to produce $\Jvec$. Tapenade is a source transformation AD tool that takes as input the source code of a mathematical function and produces a new source code that computes the derivatives of the function. Tapenade can be used to differentiate Fortran77, Fortran90, Fortran95, and C codes, with partial extensions to parallel dialects such as MPI, OpenMP, and CUDA. It has already been used to differentiate several large computational models~\cite{Gaikwad2023,maugars:hal-03759125,CARLI2023112403}.

The rest of this paper is organized as follows. Section~\ref{sec:implementation} explains the process followed for differentiating \hfbtho. Section~\ref{sec:methodology} explains our \review{testing} setup, 
and Section~\ref{sec:results} presents the results. Section~\ref{sec:conclusion} concludes the paper with a brief summary and suggestions for future work. ~\ref{section:physics} details \hfbtho's physics options that have been disabled. ~\ref{section:lapack_ad} details the external library files containing linear algebra computation that are differentiated.
~\ref{section:build_ad} presents instructions for reproducing the AD results.

\section{Implementation of \hfbthoad}
\label{sec:implementation}  
We differentiate the version of \hfbtho that was published in~\cite{perez2017axially} and was used, for example, in \cite{schunck2020calibration}. The top-level routine to differentiate is \code{HFBTHO_DFT_SOLVER}. In this initial effort we have turned off certain optional physics modules to create a simplified version of the code. These include the collective inertia package and the finite-range Gogny package. The details of these packages and other disabled options are presented in~\ref{section:physics}.

For the simplified model, the input parameters of interest are
\rhonmcode, \eoveracode, \knmcode, \asscode, \lasscode, \smasscode, \crdrcode, \cpvzcode, \crdjcode, and \cjcode since they are the parameters of the energy \review{Skyrme functional that were calibrated} on experimental measurements \cite{kortelainen2010nuclear,kortelainen2012nuclear,kortelainen2014nuclear,wild2015derivativefree,schunck2020calibration}. In this initial study we used the \hfbtho variables \code{eresu, ala2} as outputs. These arrays contain various global characteristics of the nucleus---such as the total energy and charge radius---and \review{a subset of these values} are used to evaluate the objective function (Eq.~\ref{eq:ObjectiveFunction}) in optimization-based parameter calibration. Because we have identified 14 inputs (\crdrcode, \crdjcode, \cpvzcode, \cjcode are arrays of length \code{2}) and 52 outputs (\code{eresu, ala2} are arrays of length \code{50} and \code{2}, respectively), we use the forward mode of AD, which is typically the more efficient mode when the number of outputs exceeds the number of input parameters. We additionally use vector forward mode AD, which computes the derivatives with respect to all input parameters simultaneously. 

We integrated Tapenade into the existing build system of \hfbtho by adding makefile rules to invoke Tapenade on the top-level routine. We found that Tapenade mostly worked as expected, except for two issues discussed below.

\subsection{OpenMP Shared-Memory Parallelism}
    
\hfbtho uses OpenMP loops as shown in Listing~\ref{ompcode}. Tapenade supports direct differentiation of OpenMP parallel work-sharing loops~\cite{10.1145/3472796}. Nevertheless, minor enhancements to Tapenade were required to support our syntax. The enhancements are now part of Tapenade releases.

\begin{lstlisting}[language = Fortran, rulecolor=\color{white},,label=ompcode,caption={Outline of OpenMP usage in \hfbtho}]
!$OMP Parallel Default(None)
!$OMP& SHARED(var1, var2)
!$OMP& PRIVATE(var3, var4)
       tid = OMP_GET_THREAD_NUM()
!$OMP DO SCHEDULE(DYNAMIC)
       Do ib=1,var1
!        Computation
       End Do ! ib
!$OMP End Do
!$OMP End Parallel
\end{lstlisting}

\subsection{Linear Algebra Computation}
\hfbtho uses numerous linear algebra routines provided by the BLAS and LAPACK libraries~\citep{10.1137/1.9780898719604,10.1145/567806.567807,lapackweb}. Many AD frameworks such as JAX, PyTorch, and TensorFlow, as well as tools such as Enzyme, provide customized derivatives for a subset of these routines. \review{The formulae and code in Fortran 2003 for reverse mode differentiation of BLAS matrix operations and the Cholesky factorization routine from LAPACK is given in~\citep{10.1145/3382191}. \citep{Giles2008AnEC} provides a collection of rules for the forward mode and reverse mode differentiaton of matrix operations provided by LAPACK. While deriving the forward mode formulae for BLAS and LAPACK and handcoding them is likely to result in the shortest execution times, doing so is out of the scope of this work.}  Here, as a first step, we have chosen to differentiate through the source code of the two libraries, which is possible since both libraries are written in Fortran and their source code is available online. We did so by first identifying the subset of library files that contain all the routines required by \hfbtho (see \ref{section:lapack_ad} for the list of files). Our current workflow, which we plan to improve in the future, requires all files to be concatenated into one file. Because these files are in Fortran77 format, we first preprocessed the Fortran77 files to convert the comments and line continuations in these files to match the Fortran90 format used by \hfbtho. The preprocessed files were then added to the list of files to be concatenated and subsequently differentiated.

All but one of the BLAS and LAPACK routines invoked by \hfbtho were able to be differentiated by Tapenade. The remaining routine, \code{DSYEVR}, computes selected eigenvalues and, optionally, eigenvectors of a real symmetric matrix $A$~\cite{lapackdoc}.

It may not be appropriate for an AD tool to differentiate through the source code of \code{DSYEVR} because doing so involves differentiating through a finite number of
steps in a process that is only fully resolved asymptotically. In particular, the convergence of function values (e.g., for \code{DSYEVR}) within a given tolerance does not imply the convergence of the corresponding derivative values to the same tolerance, which can lead to more inaccurate derivatives~\citep{Christianson01011994}\footnote{We note that this issue arises not just for eigensolver-based calculations, similar phenomena have been documented for basic linear system solvers \cite{more2014nd}.}. Instead, the function should be differentiated at a higher level of abstraction. 

We follow the higher-level rule for forward mode AD of \code{DSYEVR} given by~\citep{Giles2008AnEC}.
In this formulation (see Equation \ref{eqn:dsyevr}) -- which assumes that $A$ has distinct eigenvalues and real eigenvectors -- $D$ is defined to be the diagonal matrix of eigenvalues and $U$ is defined to be the matrix with columns given by the corresponding eigenvectors, so that $AU=UD$. $D$ and $U$ are computed by calling \code{DSYEVR} itself.

Denote by $\dot D$ the derivative of the eigenvalues, denote by $\dot U$ the derivative of the eigenvectors, and denote by $\dot A$ the derivative of the associated \code{DSYEVR} input. Let $X \odot Y$ denote the Hadamard (i.e., elementwise) product of two matrices 
of the same size, defined by $(X \odot Y)_{ij} = X_{ij}Y_{ij}$, and let $I$ denote an identity matrix. 
We may then compute $\dot D$ and $\dot U$ via 
\begin{align}
\label{eqn:dsyevr}
  & \dot D = I \odot (U^{-1} \dot A U) \nonumber \\
  & \dot U =  U ( F \odot (U^{-1} \dot A U)), \\
& \mbox{where $F$ is defined elementwise by } F_{ij} =
    \begin{cases}
      (d_j-d_i)^{-1} & \mbox{for } i \ne j \nonumber\\
      0 & \text{otherwise.}      \nonumber \\
    \end{cases}                 \nonumber   
\end{align}     

 We have thus separately coded the forward mode derivatives for \code{DSYEVR} and used that code in place of the code generated by Tapenade. 

\section{\review{Testing} Methodology}\label{sec:methodology}
Our \review{tests} serve to 1) verify that the derivatives computed by different configurations are consistent and that we are able to reproduce historical results using the Jacobians produced by HFBTHO-AD, and 2) compare the performance of AD and finite-difference approximation.

\subsection{Comparison Parameter Points}
For validation purposes, we have chosen three different parameter points from the UNEDF0 study~\cite{kortelainen2010nuclear}. The first is the \slyf parameterization~\cite{EChabanat_1995}, which was used as the starting point for the optimization-based parameter calibrations.
The second, \unedfnb, is the UNEDF0 local minimizer determined via an optimization that imposed no bound constraints on parameter values,  and was reported in Table VII of
\cite{UNEDF0}. The third point,  \unedfpre, is the UNEDF0 local minimizer reported in Table VIII of \cite{UNEDF0} that was determined via a bound-constrained optimization designed to ensure that the minimizer had parameter values within physically acceptable ranges.  For this solution, both \smasscode and \knmcode are at a bound constraint.

\begin{table}
\center
\begin{tabular}{ll}
\hline
 {\bf Intel/LAPACK} & \multirowcell{8}[0pt][l]{Intel OneAPI v2023.2.1 w/OpenMP \\
 HDF5 v1.14.3 \\
 Intel OneAPI MPI v2021.10.0} \\
 Netlib-LAPACK v3.12.1  \\
 \cline{1-1}
 {\bf Intel/MKL}\\
 Intel OneAPI MKL v2023.2.0  \\
\cline{1-1}
 {\bf Intel/LAPACK/AD} \\
 Netlib-LAPACK v3.12.1  \\
 OpenJDK v17.0.11\_9 \\
 Tapenade (commit d0c20bda)\\
 \hline
  {\bf GCC/LAPACK/AD} \\
  Netlib-LAPACK v3.9.0  & GCC v13.2.0 w/OpenMP \\
  OpenJDK v17.0.11\_9 & HDF5 v1.14.3 \\
  Tapenade (commit d0c20bda)& MPICH v4.2.0\\
\hline
\end{tabular}
\caption{Software configurations \label{tab:configurations}}
\end{table}

\subsection{Software Architecture}
Our software architecture consists of an \hfbtho executable at an \emph{inner level} that computes the model functions $m_i$.  Each invocation of the executable provides all values computed
for a given nucleus configuration at a given $\xvec$.  A single evaluation of
$\rvec$ requires running the \hfbtho executable for the $\nn=72$ nucleus configurations associated with
the dataset. A Python package was written at an \emph{outer level} to evaluate both $\rvec$ and $\Jvec$ for AD-enabled \hfbtho using MPI
parallelization to take advantage of the trivial parallelization across the 72 computational tasks.

\subsection{Testing Environment} 
All data acquisition and \review{tests} were run on multiple nodes of the Improv cluster
managed by the Laboratory Computing Resource Center of Argonne National Laboratory~\cite{improvweb}.  Each node
contains two sockets each holding a single AMD EPYC 7713 64-core processor, and
each core uses a single hardware thread. 

\subsubsection{Inner-Level Software Configurations} We have constructed multiple software configurations for the inner level as shown in 
Table~\ref{tab:configurations}.  The Intel/LAPACK configuration is used for data acquisition tasks that only need to evaluate $f$ and $\rvec$ (e.g.,
\ecnoise and forward differences). The Intel/MKL configuration, which was used for derivative-free optimization and sensitivity analysis in previous studies because it was determined to be resource efficient, is used here only for confirming software robustness of the Intel/LAPACK configuration.
The Intel/LAPACK/AD configuration is used to evaluate $f, \rvec$, and $\Jvec$. This configuration exclusively uses the open-source Netlib LAPACK because Tapenade requires access to the source code in order to differentiate it.
The GCC/LAPACK/AD configuration is used for confirming robustness of our builds and verifying the output of the Intel/LAPACK/AD configuration.

The Intel/LAPACK/AD and GCC/LAPACK/AD configurations are designed to give the derivatives with respect to the 14 input parameters used in the UNEDF2 configuration as explained in Section~\ref{sec:implementation}. Because our comparison parameter points are from the UNEDF0 study, however, the finite differences compute the derivatives only with respect to the 12 parameters calibrated in that study.

For all non-AD configurations, the OpenMP stack size is set to 16 MB.
For all AD configurations, the OpenMP stack size is set to 10 GB to accommodate the memory needs of
the parallelized sections of the Tapenade-generated differentiation code.
~\ref{section:build_ad} presents the full details of generating and executing the Intel/LAPACK/AD configuration.

\subsubsection{OpenMP Threads per MPI Process} 
\label{sec:omp}
\begin{figure}[t]
\includegraphics[width=5.5in]{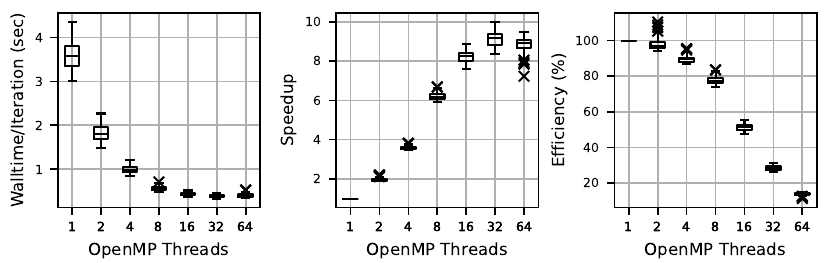}\\
\includegraphics[width=5.5in]{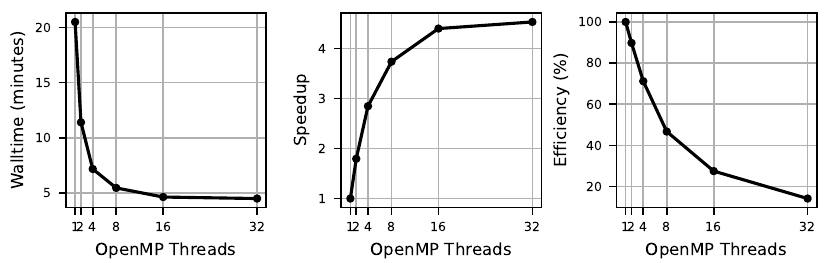}
\caption{Strong-scaling results for the (top) Intel/LAPACK software configuration
and the (bottom) Intel/LAPACK/AD software configuration.
\label{fig:fdadruntime}}
\end{figure}

Strong-scaling tests were run to determine how many OpenMP threads to use per
MPI process and how many nodes to run per job.  For setting up the non-AD test
environment, the residual function $\rvec$ was evaluated by using the full software
architecture based on Intel/LAPACK with different numbers of OpenMP threads per
MPI process at \unedfpre.  For each thread count setup, the walltime needed to run each of
the 72 \hfbtho computations was divided by the number of iterations run by
\hfbtho to converge to the solution of the computation.  This approach allows for direct
comparison of the walltime per iteration across all \hfbtho computations, which is
shown in the boxplot visualization in Figure~\ref{fig:fdadruntime}.  The figure
also shows the speedup and parallel efficiency derived from the data.  
We note that this
design assumes that the scaling results at this single point are representative
of scaling across the parameter space.

Because AD-based \hfbtho computations \review{involve significantly longer walltimes}, a simpler version of the
scaling test was performed for setting up the AD test environment.  In
particular, the single \hfbtho computation for computing observables of the $^{160}$Dy deformed ground-state configuration at the \unedfpre solution was run
directly with the Intel/LAPACK/AD \hfbtho executable with different numbers of
OpenMP threads.  Therefore, this test assumes that the scaling results,
displayed in Figure~\ref{fig:fdadruntime}, for this single computation are representative of
scaling for the full set of computations across the parameter space.

The scaling results in Figure~\ref{fig:fdadruntime} are typical of an
application whose performance does not scale close to ideal scaling.  The number of OpenMP
threads per MPI process for non-AD evaluations (8 threads/process) and for AD-based
evaluations (4 threads/process) was chosen as a balance between decreasing
walltimes and making efficient use of the hardware.  In addition, these
choices correspond, respectively, to similar parallel efficiencies of 78\% and
71\%, which should improve the ability to compare performance results cleanly. Our choices also emphasize that \review{there can} be additional tradeoffs to consider -- based on resource availability, time and energy constraints, etc. -- when deploying AD-based solutions.

\subsubsection{MPI Processes per Job} 
\label{sec:mpi}
The walltime needed to perform a single computation can vary strongly from one
nucleus configuration to the next as well as from one parameter point to the next.  As a
result, evaluating $\rvec$ at a single point with the number of MPI processes
set to the number of underlying \hfbtho computations would likely result in poor
resource usage due to MPI processes idling.  Therefore, significantly fewer MPI
processes are typically used to (hopefully) achieve better load balancing and hence better resource usage.  Therefore, all non-AD jobs are run using 32
MPI processes across two nodes; for AD-based jobs, 32 MPI processes on one node.
This matching of MPI process count ensures a similar load-balancing strategy,
which should also improve the ability to compare cleanly performance results
such as CPU usage or core-hour resource utilization.

\subsection{Noise Levels and Difference Parameters}
\label{sec:differenceparameters}
We employ \ecnoise~\cite{more2011ecn} to obtain independent scalar magnitude estimates of the deterministic computational noise, $\epsilon_{ij}$, present in $r_i$, restricted to the $j^{\rm th}$ parameter, at a fixed input $\xvec_0$. 
Using a technique outlined in \cite{more2011edn} that also coarsely approximates
the second derivative of $r_i$ with respect to the $j^{\rm th}$ parameter, we obtain
individually for each $(i, j)$ pair
the forward difference parameter $h_{ij}$ using $\epsilon_{ij}$. 
This specific choice of $h_{ij}$ is generally found in practice to yield
reliable FD approximations, and theoretical results concerning its resulting
approximation error have been established under particular assumptions on the nature of deterministic noise \cite{more2011edn}. 
In general, computing all
$h_{ij}$ at a single $\xvec_0$ via this enumerative approach requires a constant (on the order of $10$) times the dimension ($\np$) number of evaluations of $\rvec$ in a neighborhood about $\xvec_0$.
We, therefore, expect to see the AD-based approach perform better than the FD-based approach (which includes the determination of $h_{ij}$), underscoring the former's value, especially as $\np$ grows. 

\begin{figure}[t!]
\begin{center}
\includegraphics[width=5.5in]{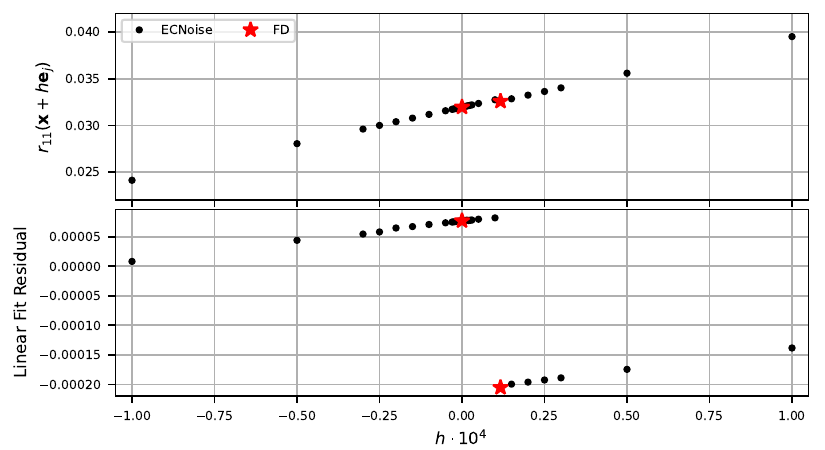}
\end{center}
\caption{Noise level determined for the $r_{11}$ residual associated with the ground-state binding energy of $^{248}$Fm
at \unedfpre along \smasscode ($\evec_j$). The noise level  is abnormally large, and there
is a large deviation between the partial derivative of $r_{11}$ with respect to
\smasscode approximated with AD (78.45) and with forward differences (53.55).
(top) \ecnoise data collected at \unedfpre to study the noise level.  The
two FD evaluations used to approximate the partial derivative of this observable with
respect to \smasscode are shown in red. (bottom) The deviation between the
\ecnoise data and the predictions made by a linear regression model that was fit to the
\ecnoise data with least-squares.  There is a  sharp change in the \ecnoise data and the two
evaluations used to compute the forward difference lie on either side of this change.}
\label{fig:HfbthoJumpExample}
\end{figure}

\ecnoise results gathered as part of determining $h_{ij}$ parameters revealed
that in some cases \hfbtho results in computational noise that could
limit the use of \ecnoise for determining $h_{ij}$, whether in difficulties in obtaining accurate
$h_{ij}$ or subsequent accurate finite-difference approximations.  Figure~\ref{fig:HfbthoJumpExample} shows, for example, that the two
function evaluations needed to approximate a derivative with FD can lie on either side of a quick and strong change in function values that is likely a numerical
artifact (e.g., associated with a differing number of internal iterations to achieve a prescribed tolerance) rather than a true change in the underlying function.  

Fully understanding this numerical issue (see, e.g., \cite[Figure 3.2 right]{more2014nd}) and adapting
the use of \ecnoise appropriately is outside the scope of this work. We instead set an
internal \ecnoise tolerance sufficiently high such that all $h_{ij}$ are determined without failures. While some of the $h_{ij}$ are therefore potentially inaccurate, the agreement between AD and FD-based Jacobian matrices is still reasonably good, as shown in subsequent sections. 

\section{\review{Results}}
\label{sec:results}
In this section we present validation and performance results.

\subsection{Output Comparison}

\begin{figure}
\begin{center}
\includegraphics[width=5.5in]{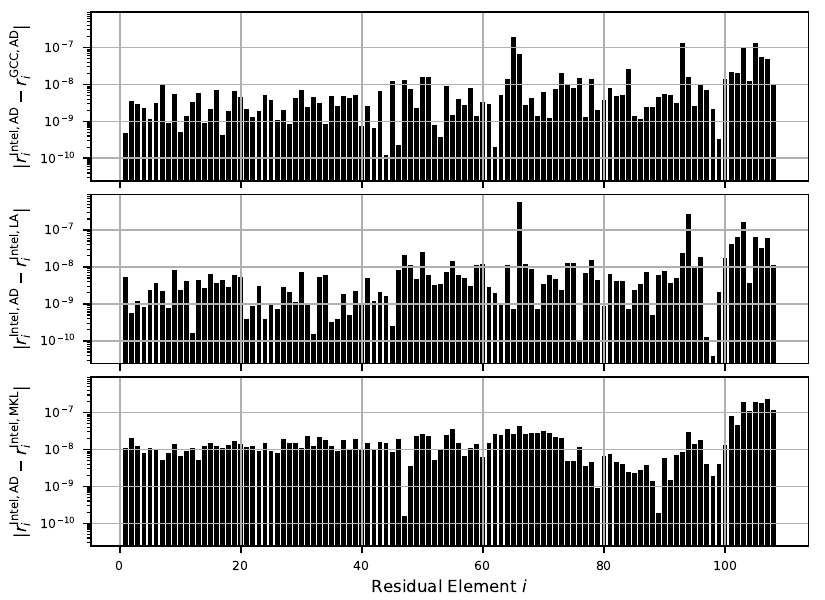}
\end{center}
\caption{Absolute deviations between the residual vectors $\rvec$, obtained at \unedfpre
as computed using (top) Intel/LAPACK/AD and GCC/LAPACK/AD, (middle)
Intel/LAPACK/AD and Intel/LAPACK, and (bottom) Intel/LAPACK/AD and Intel/MKL.}
\label{fig:PrimalComparison}
\end{figure}

In Figure~\ref{fig:PrimalComparison} we compare the deviations in $\rvec$ computed by
Intel/LAPACK, Intel/LAPACK/AD, and GCC/LAPACK/AD at \unedfpre. The deviations are all below $10^{-6}$ and indicate that applying Tapenade did not introduce any considerable errors in the residual computation. Similar results are obtained at \unedfnb and \slyf. 

\subsection{Validation of Derivatives}

\begin{figure}[p]
\includegraphics[width=5.5in]{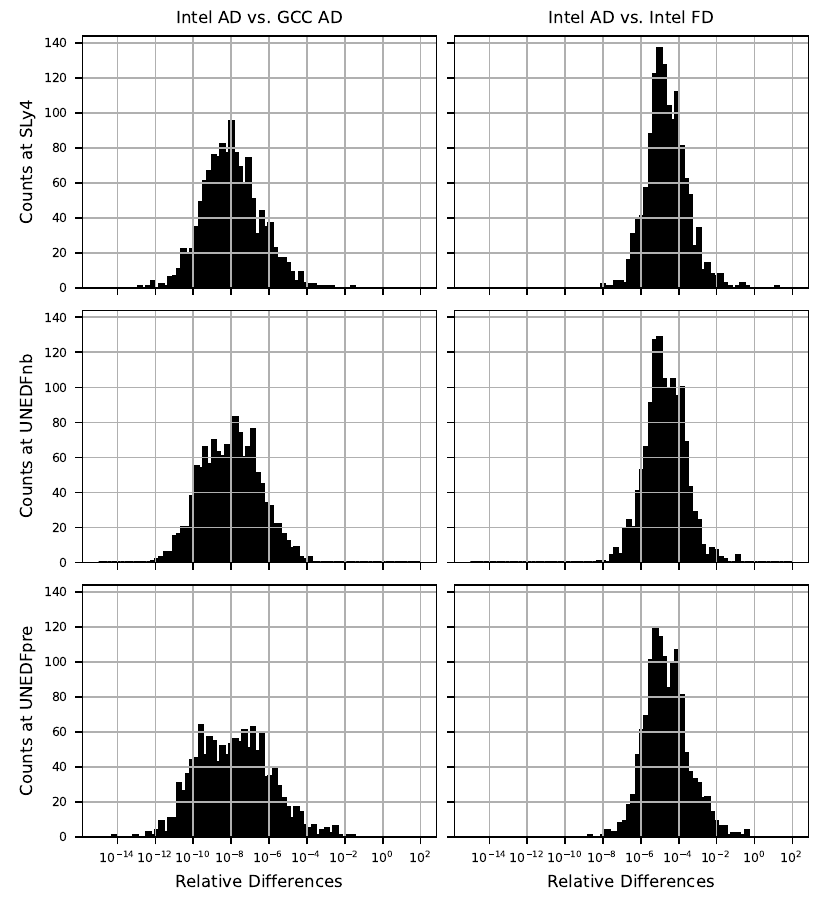}
\caption{Histograms of relative deviations of all partial derivatives in the Jacobian matrices obtained at the three comparison parameter points between (left) Intel/LAPACK/AD and GCC/LAPACK/AD and (right) between Intel/LAPACK/AD and FD using Intel/LAPACK.  All relative deviations were computed as $\left|1.0-{J_{\textrm{other}_{ij}}}/{J_{\textrm{reference}_{ij}}}\right|$ using the Intel/LAPACK/AD as the reference value.
\label{fig:adresults}}
\end{figure}

We now compare the Intel/LAPACK/AD and GCC/LAPACK/AD Jacobian matrices computed at \unedfpre, \unedfnb, and \slyf. At each point we compare the relative deviation between the $108 \times 12 = 1296$ elements of $\Jvec$ computed by the two configurations. Figure~\ref{fig:adresults} (left column) shows that computed derivatives of the two configurations match  well at the three parameter points. This result  indicates that our approach is robust.

Next we compared the values of $\Jvec$ 
computed by
Intel/LAPACK/AD and FD using Intel/LAPACK
at the three parameter points. The $h_{ij}$ for FD are determined by the approach explained in  Section~\ref{sec:differenceparameters}. Figure~\ref{fig:adresults} (right column) shows that computed derivatives of the two configurations match well. There are, however, a small number of elements of $\Jvec$ in each case where the relative deviation is greater than $10^{-2}$. The mismatch is likely due to the presence of abnormal computational noise, as highlighted in Section~\ref{sec:differenceparameters}.

\begin{figure}
\includegraphics[width=5.5in]{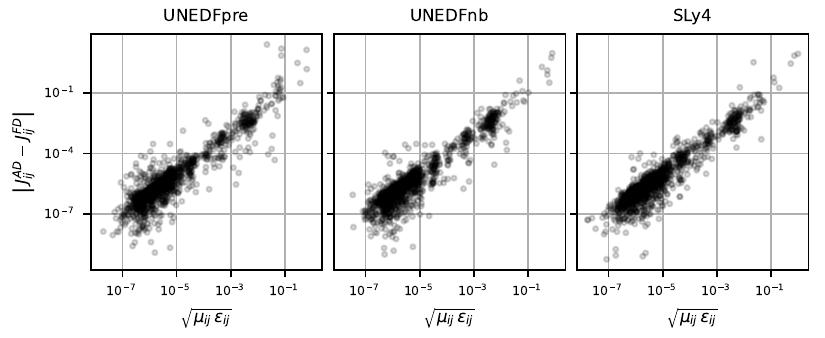}
\caption{Comparison of the absolute deviations between the Intel/LAPACK/AD Jacobian and the Intel/LAPACK/FD Jacobian approximation to a theoretical error approximation~\cite{more2011edn} derived from the \ecnoise intermediate quantities $\epsilon_{ij}$ and $\mu_{ij}$. 
\label{fig:ErrorScaling}}
\end{figure}

In \Cref{fig:ErrorScaling}, we plot the absolute deviations in the entries of Intel/LAPACK/AD and Intel/LAPACK/FD ($|J_{ij}^{AD} - J_{ij}^{FD}|$) against the square root of the product of $\epsilon_{ij}$ with a quantity $\mu_{ij}$. 
This quantity $\mu_{ij}$ is the previously mentioned coarse estimate of the second derivative $\partial_{jj} r_i$ that is computed as part of the \ecnoise procedure for determining $h_{ij}$. Although the conditions of \cite[Theorem 3.2]{more2011edn} are not precisely satisfied, this result suggests that the true Jacobian $\Jvec$ ought to satisfy, up to multiplicative constants, $|J_{ij} - J_{ij}^{FD}| \approx \sqrt{\epsilon_{ij}\mu_{ij}}$. 
The trend in \Cref{fig:ErrorScaling} suggests that $|J_{ij}^{AD} - J_{ij}^{FD}| \approx \sqrt{\epsilon_{ij}\mu_{ij}}$, lending further credence to the claim that $\Jvec^{AD}$ is a good approximation to $\Jvec$.

Lastly, we examine the computed derivatives when used for sensitivity analysis. At each optimization solution $\xHat$, the UNEDF0 study used second-order central
differences to approximate the Jacobian matrix $\Jvec(\xHat)$.  The matrix was
used to compute different quantities such as the statistical uncertainties
associated with each coordinate in the solution, which are the diagonal elements
of the covariance matrix \cite{Seber1989},
\[
\frac{f(\xHat)}{\np - d}\left(\Jvec(\xHat)^T \Jvec(\xHat)\right)^{-1}.
\]
Table~\ref{tab:StdDevComparison} presents a modified version of the
uncertainties published in \cite{kortelainen2010nuclear} along with the same uncertainties
computed using the Jacobian matrices reported here.  
Despite \hfbtho undergoing several improvements since the original UNEDF0 study and determining good forward and central difference parameters being inherently difficult, the new results reproduce the original results well.

\begin{table}
\begin{center}
\begin{tabular}{|l|c|r|r|r|}
\hline
Parameter & Name & $\sigma_{AD}$ & $\sigma_{FD}$ & $\sigma$ \\
\hline
\rhonm  & \rhonmcode   &  0.0005 &  0.0005 &  0.0008 \\
\eovera & \eoveracode  &  0.046  &  0.046  &  0.046  \\
\ass    & \asscode     &  2.554  &  2.555  &  2.561  \\
\lass   & \lasscode    & 33.417  & 33.456  & 33.525  \\
\crdrz  & \crdrcode(0) &  1.418  &  1.418  &  1.421  \\
\crdro  & \crdrcode(1) & 47.529  & 47.622  & 47.699  \\
\cpvzz  & \cpvzcode(0) &  1.760  &  1.758  &  1.763  \\
\cpvzo  & \cpvzcode(1) &  2.782  &  2.783  &  2.806  \\
\crdjz  & \crdjcode(0) &  2.860  &  2.860  &  2.866  \\
\crdjo  & \crdjcode(1) & 24.613  & 24.613  & 24.668  \\
\hline
\end{tabular}
\end{center}
\caption{Comparison of the statistical uncertainties of the \unedfpre solution
derived using the AD-based Jacobian matrix ($\sigma_{AD}$), the FD-based
Jacobian matrix ($\sigma_{FD}$), and the original uncertainties. Because \smasscode and \knmcode are actively bound at \unedfpre, these are not included in this table. We believe
that the original uncertainties reported in Table VIII in \cite{kortelainen2010nuclear}
correspond to the diagonal elements of $\left(\Jvec^T \Jvec\right)^{-1}$ rather than of
the covariance matrix.  The reference values $\sigma$ reported here have been
corrected accordingly using the values in Table VIII and the $f(\xHat)$ result
of the original study. \label{tab:StdDevComparison}}
\end{table}

\subsection{Performance Analysis}

\begin{table}[t]
\begin{center}
\begin{tabular}{|l|r|r|r|r|r|}
\hline
$\mathbf{x}$ & $T_{\rvec}$ & $T_{\mathrm{FD}}$ & $T_{\mathrm{AD}}$ & $T_{\mathrm{FD}}/T_{\rvec}$ & $T_{\mathrm{AD}}/T_{\rvec}$ \\
\hline
\slyf     & 0.70 & 10.77 & 12.20 & 15.50 & 17.55 \\
\unedfnb  & 0.85 & 13.00 & 15.53 & 15.24 & 18.20 \\
\unedfpre & 1.12 & 17.60 & 20.12 & 15.67 & 17.91 \\
\hline
\end{tabular}
\end{center}
\caption{Walltimes (in minutes) for evaluating the residual function
$\rvec(\xvec)$ using Intel/LAPACK ($T_{\rvec}$), evaluating $\rvec(\xvec)$ using Intel/LAPACK,
and approximating $\Jvec(\xvec)$ using Intel/LAPACK with forward differences given a set of $h_{ij}$
($T_{\mathrm{FD}}$), and simultaneously evaluating $\rvec(\xvec)$ while
approximating $\Jvec(\xvec)$ with using Intel/LAPACK/AD ($T_{\mathrm{AD}}$).
The latter two columns provide the ratio of the (approximate) derivative evaluation walltime to the residual evaluation walltime.
\label{tab:WalltimeComparison}}
\end{table}

\begin{table}
\begin{center}
\begin{tabular}{|l|r|r|r|r|r|}
\hline
$\mathbf{x}$ & $U_{\rvec}$ & $U_{\mathrm{FD}}$ & $U_{\mathrm{AD}}$ & $U_{\mathrm{FD}}/U_{\rvec}$ & $U_{\mathrm{AD}}/U_{\rvec}$ \\
\hline
\slyf     & 2.97 & 45.95 & 26.03 & 15.50 & 8.78 \\
\unedfnb  & 3.64 & 55.47 & 33.13 & 15.24 & 9.10 \\
\unedfpre & 4.79 & 75.09 & 42.92 & 15.67 & 8.95 \\
\hline
\end{tabular}
\end{center}
\caption{Resource utilization in core-hours needed for evaluating the
residual function $\rvec(\xvec)$ without AD ($U_{\rvec}$), evaluating
$\rvec(\xvec)$ without AD and approximating $\Jvec(\xvec)$ with forward differences
given a set of $h_{ij}$ ($U_{\mathrm{FD}}$) \review{with respect to 12 parameters}, and simultaneously evaluating
$\rvec(\xvec)$ while approximating $\Jvec(\xvec)$ with AD ($U_{\mathrm{AD}}$) \review{with respect to 14 parameters}.
\label{tab:CorehoursComparison}}
\end{table}

Tables~\ref{tab:WalltimeComparison} and \ref{tab:CorehoursComparison} present performance data for evaluating the objective and Jacobian functions at different parameter points using the two principal software configurations, Intel/LAPACK and Intel/LAPACK/AD, used in this study. The number of MPI processes and OpenMP threads used to gather performance data for these configurations is based on the analysis in Sections~\ref{sec:omp}~and~\ref{sec:mpi}. For approximating Jacobian matrices using FD, we do not include the time taken to use \ecnoise to determine appropriate $h_{ij}$ values.  Because of the present implementation of AD in \hfbtho, AD-based performance results are penalized for the computation of derivatives with respect to 14 parameters rather than for only the 12 parameters that are free for UNEDF0. 

Table~\ref{tab:WalltimeComparison} presents the walltimes required for function evaluations.  The average FD runtime ratio of $T_{\mathrm{FD}}/T_{\rvec} = 15.47$ is larger than the $(\np+1=) 13$ that we would expect, which could be due to \hfbtho computations $r_i(\xvec + h_{ij}\evec_j)$ that require more iterations than those needed to compute $r_i(\xvec)$.  Based on prior experience, the FD walltimes might be decreased by using, for example, the Intel/MKL software configuration.

The average AD runtime ratio of $17.89$ for computing $\rvec$ as well as $\Jvec$ with respect to 14 parameters using vector forward mode AD is within the theoretical upper bound expressed in ~\cite[Section 4.5]{Griewank2008EDP}. 
The exact runtime ratio depends on the number of memory accesses and number of multiplication operations compared with additions and nonlinear intrinsics. Better performance may be achieved by differentiating with respect to only the 12 parameters that affect the output in UNEDF0.

Since the MPI and OpenMP setups result in the two software configurations using different amounts of hardware to achieve similar load balancing with similar parallel efficiency, Table~\ref{tab:CorehoursComparison} presents resource allocation quantified as core-hours to aid direct comparison \review{of resource usage by the two differentiation approaches}.  Evaluating $\rvec$ and approximating $\Jvec$ with FD requires on average $15.47$ times the resource allocation of evaluating just $\rvec$. Simultaneously evaluating $\rvec$ and approximating $\Jvec$ with Intel/LAPACK/AD requires $8.94$ times the resources required for evaluating $\rvec$ with Intel/LAPACK.
\review{Note that the FD-based resource requirement ratios are equal to the FD-based runtime ratios because both $T_{\rvec}$ and $T_{\mathrm{FD}}$ were acquired using the same Intel/LAPACK hardware setup, which used two nodes.  However, the AD-based resource requirement ratios are half that of its runtime ratios because $T_{\mathrm{AD}}$ was acquired using the Intel/LAPACK/AD hardware setup, which used only one node.  Therefore, while the FD differentiation technique is superior in terms of walltime for these test setups, the AD-based differentiation technique is superior in terms of resource usage since less hardware is used to achieve the same parallel efficiency as for the FD-based technique.}

\section{Conclusion and Future Work}
\label{sec:conclusion}
We have used the AD tool Tapenade to differentiate the EDF solver \hfbtho. 
We have differentiated through the BLAS and LAPACK routines that are invoked by \hfbtho (except for \code{DSYEVR}). We have computed derivatives at three parameter points from the UNEDF0 study and validated them using finite differences and successfully reproduced statistical uncertainties of the \unedfpre solution.
\review{Our performance comparison shows that within the context of our particular test environments the AD and FD differentiation techniques have similar walltimes but that AD performance is superior in terms of resource usage.}
However, FD also requires judicious determination of difference parameters, which adds additional overhead to FD-based approximation and is a key reason for favoring AD over FD.

In future work we will employ the computed derivatives to supply Jacobians to a variation on a Levenberg--Marquardt optimization method for nonlinear least squares, and we will compare its performance with that of prior derivative-free optimization approaches.
We will further study \hfbtho's noisy behavior, with the aim of  understanding and mitigating the abnormal noise discussed in Section~\ref{sec:differenceparameters} and shown in Figure~\ref{fig:HfbthoJumpExample}, to inform the optimization algorithm. We also will differentiate the latest release of \hfbtho. This will enable us to perform new studies of nuclear energy functionals -- for example, a comparison of Skyrme and Gogny functionals calibrated on the same dataset. 
This will also open the way to performing EDF calibration and uncertainty quantification while restoring broken symmetries, a process that we presently believe is computationally too expensive for derivative-free methods.

\section*{Acknowledgment}
This work was supported in part by the U.S.\
Department of Energy, Office of Science, Office
of Advanced Scientific Computing Research Applied Mathematics and SciDAC programs, and Office of Nuclear Physics SciDAC program under Contract
Nos.\ DE-AC02-06CH11357, DE-AC02-05CH11231, and DE-AC52-07NA27344. 
This work was partially performed under the auspices of the U.S.\ Department of Energy by Lawrence Livermore National Laboratory under contract DE-AC52-07NA27344, managed by Lawrence Livermore National Security, LLC.
We gratefully acknowledge the computing resources provided on Improv and Bebop, high-performance computing clusters operated by the Laboratory Computing Resource Center at Argonne National Laboratory. We thank Paul Hovland, Jan H\"uckelheim, Rodrigo Navarro P\'{e}rez, and Lars Zurek.

\appendix

\section{Physics Modules and Options That Are Disabled}
\label{section:physics}

The following is a list of modules that are disabled in order to simplify the code for AD. These modules are not relevant in the optimization of the parameters of Skyrme-like energy functionals.
\begin{itemize}
  \item \code{Collective inertia package}: calculates the collective inertia mass tensor at the perturbative cranking approximation. This package is  relevant only in fission calculations.
  \item \code{Finite-range Gogny package}: computes the matrix elements and expectation value of the finite-range Gogny force. This package is inactive when working with Skyrme-like energy functionals.
  \item \code{Particle number projection module}: restores particle number by employing projection techniques. This option is not currently used when performing the calibration of energy functionals.
  \item \code{Transformed harmonic oscillator basis package}: performs a local transformation of harmonic oscillator basis functions to improve the asymptotic behavior at large distances. This package is  relevant only for describing the physics of weakly bound nuclei.
\end{itemize}

The following is a list of Makefile preprocessor options that are also disabled in order to simplify the code for AD. They activate/deactivate sections of the code that are not relevant for the optimization of Skyrme-like energy functionals.

\begin{itemize}
  \item \code{USE_MPI}: enables MPI. Both in direct optimization or Bayesian calibration of the energy functional, \hfbtho runs on a single node with shared-memory parallelism only, making this option inactive.
  \item \code{DRIP_LINES}: when enabled,  calculates an entire mass table from the proton to the neutron dripline. This option is  used only in production runs and is not relevant for the calibration of energy functionals.
  \item  \code{DO_MASSTABLE}: when enabled, calculates a section of the mass table. Like \code{DRIP_LINES}, this option is  designed only for production runs and is not relevant for the calibration of energy functionals.
  \item  \code{DO_PES}: when enabled,  calculates a potential energy surface for a set of nuclei. This option is mostly used in large-scale calculations of fission and is not relevant in the calibration of energy functionals.
  \item  \code{READ_FUNCTIONAL}: when enabled, reads the parameters of the functional from a file. Otherwise, the energy functional is defined in the code based on the value of the input keyword ``functional.'' In the optimization of Skyrme-like energy functionals, the parameters of the functional are passed directly to \hfbtho through an external wrapper, making this option irrelevant.

  \item  \code{USE_LOCALIZATION}: when enabled,  computes the nuclear localization functions. This is  relevant only for production runs.
  \item  \code{GOGNY_SYMMETRIES}:  assumes several symmetries in the finite-range matrix elements defining the Gogny force. This option is not relevant when working with Skyrme-like energy functionals.
  \item  \code{GOGNY_HYPER}: when enabled,  uses hypergeometric function to calculate the matrix elements of the finite-range component of the Gogny force. It is not relevant for work with Skyrme-like energy functionals
  \item  \code{USE_QRPA}: when enabled,  produces output for the QRPA-pnFAM code. This option is not relevant for calculations performed at the HFB level.
\end{itemize}

\section{Linear Algebra Files}
\label{section:lapack_ad}
The linear algebra files used by \hfbtho from LAPACK \verb|v3.12.1| are given in \Cref{lapack_files}, \Cref{blas_files}, \Cref{install_files}, and \Cref{handcoded_files}. The files in \Cref{handcoded_files} cannot be differentiated as is. They require  changes to the code  because of limitations of Tapenade, because of the way the code is used in \hfbtho, or because differentiating through the function is not appropriate.

\begin{center}
\begin{table}[t]
\caption{\label{lapack_files} LAPACK Files from \code{./SRC/}}
\begin{tabular}{|l|l|l|l|l|l|}
\hline
dlansy.f&
dsytrd.f&
dsterf.f&
dstemr.f&
dormtr.f&
dstebz.f\\
dstein.f&
dlasyf.f&
dsytf2.f&
dgetrf2.f&
dlaswp.f&
dsyevr.f\\
dsyevx.f&
dsytrf.f&
dpotrf.f&
dsytri.f&
dpotri.f&
dgetrf.f\\
dgetri.f&
dtrtri.f&
dlagtf.f&
dlagts.f&
disnan.f&
dlascl.f\\
dtrti2.f&
dlacpy.f&
dlarrc.f&
dlae2.f&
dlaebz.f&
dlaev2.f\\
dlaisnan.f&
dlaneg.f&
dlanst.f&
dlapy2.f&
dlar1v.f&
dlarf.f\\
dlarfb.f&
dlarfg.f&
dlarft.f&
dlarnv.f&
dlarrv.f&
dlarra.f\\
dlarrb.f&
dlarrd.f&
dlarre.f&
dlarrf.f&
dlarrj.f&
dlarrk.f\\
dlarrr.f&
dlaruv.f&
dlaset.f&
dlasq2.f&
dlasq3.f&
dlasq4.f\\
dlasq5.f&
dlasq6.f&
dlasrt.f&
dlatrd.f&
dorm2l.f&
dorm2r.f\\
dormql.f&
dormqr.f&
dsytd2.f&
ieeeck.f&
iladlc.f&
iladlr.f\\
ilaenv.f&
iparmq.f&
xerbla.f&
dcombssq.f&
dlarf1f.f&
dlarf1f.f\\
dlarf1l.f&
dlassq.f90&&&&\\
\hline
\end{tabular}
\end{table}

\begin{table}
\caption{\label{blas_files} BLAS Files from \code{./SRC/BLAS/}}
\begin{tabular}{|l|l|l|l|l|l|}
\hline
ddot.f&
dswap.f&
dscal.f&
dcopy.f&
daxpy.f&
dgemm.f\\
dgemv.f&
dtrsm.f&
dsymv.f&
idamax.f&
dsyr.f&
dsyr2.f\\
dsyr2k.f&
dtrmm.f&
dtrmv.f&
dger.f&
dgemmtr.f&
dnrm2.f90\\
\hline
\end{tabular}
\end{table}

\begin{table}[h!]
\caption{\label{install_files} Install Files from \code{./INSTALL/}}
\begin{tabular}{|l|l|}
\hline
lsame.f & dlamch.f \\
\hline
\end{tabular}
\end{table}

\begin{table}[h!]
\caption{\label{handcoded_files} Handcoded Files}
\begin{tabular}{|l|l|}
\hline
dlagts.f& Rewrite to remove use of three arguments {\tt MAX}\\
dlanst.f& \\
dlarfg.f& Add precision from \hfbtho\\
dsyevr.f& Differentiated as explained in Section  \ref{sec:implementation} \\
\hline
\end{tabular}
\end{table}
\end{center}

\section{Reproducing Results}
\label{section:build_ad}

\noindent
Provided here is a brief description of creating the inner-level INTEL/LAPACK/AD configuration and executing it. The GCC/LAPACK/AD configuration requires a similar process. 

\subsection{Obtaining \hfbtho UNEDF0}
\label{subsection:hfbtho}

\noindent We have created a snapshot of \hfbtho at \url{https://doi.org/10.5281/zenodo.16249941},  to ensure reproducibility. Download the file \code{hfbtho.tar.gz} and then unpack it.

\begin{lstlisting}[language = bash, rulecolor=\color{white}]
$ tar zxvf hfbtho.tar.gz
\end{lstlisting}

\subsection{Installing Tapenade}
\label{subsubsection:install_tapenade}

\noindent
    We detail below the instructions for Linux, but the latest instructions for many operating systems can be found at \url{https://tapenade.gitlabpages.inria.fr/tapenade/distrib/README.html}.

\begin{enumerate}

\item Clone the code

\begin{lstlisting}[language = bash, rulecolor=\color{white}]
$ cd; git clone https://gitlab.inria.fr/tapenade/tapenade.git
$ cd tapenade
\end{lstlisting}

\item Use the version of Tapenade from Table~\ref{tab:configurations}. This step is optional. As underlying dependencies of Tapenade change, it may be easier to build the latest version of Tapenade.
\begin{lstlisting}[language = bash, rulecolor=\color{white}]
$ git checkout 
d0c20bdae18ad1c12e2578103cc28ce5e4995570
\end{lstlisting}

\item On Linux, depending on your distribution, Tapenade may require you to set the shell variable \verb|JAVA_HOME| to your Java installation directory. It is often \verb|JAVA_HOME=/usr/java/default|. You might also need to modify the \verb|PATH| by adding the bin directory from the Tapenade installation. Every time you wish to use the AD capability with Tapenade, you must first ensure that this Java environment is loaded and that all necessary environment variables are properly set. We have used \verb|OpenJDK 17.0.11_9| here to build Tapenade.

\item Build Tapenade
\begin{lstlisting}[language = bash, rulecolor=\color{white}]
$ ./gradlew clean ; ./gradlew
\end{lstlisting}
\end{enumerate}

You should now have a working copy of Tapenade. For more information on the \code{tapenade} command and its arguments, type the following.

\begin{lstlisting}[language = bash, rulecolor=\color{white}]
$ tapenade -?
\end{lstlisting}

\subsection{Installing LAPACK}
We detail below the instructions for building and installing LAPACK \verb|v3.12.1| with Intel's \verb|ifort| compiler. The instructions for using other compilers or LAPACK \verb|v3.9.0| are similar.
\begin{itemize}
\item Obtain LAPACK

\begin{lstlisting}[language = bash, rulecolor=\color{white}]
$ cd; wget https://github.com/Reference-LAPACK/lapack/archive/v3.12.1.tar.gz
$ tar zxvf v3.12.1.tar.gz
\end{lstlisting}

\item Build LAPACK

\begin{lstlisting}[language = bash, rulecolor=\color{white}]
$ cd lapack-3.12.1
$ cp INSTALL/make.inc.ifort ./make.inc
$ make
\end{lstlisting}

\item Build the additional target \verb|la_xisnan.mod| 

\begin{lstlisting}[language = bash, rulecolor=\color{white}]
$ cd SRC
$ rm la_xisnan.o
$ make la_xisnan.o
\end{lstlisting}
\end{itemize}

If the build products are moved to a separate installation location, make certain to include the manually built \texttt{la\_xisnan.mod} file in the installation.

\subsection{Other Dependencies}
The dependencies listed below must be installed.
\begin{enumerate}
 \item \verb|Python 3.X|
 \item \verb|HDF5|
\end{enumerate}

In case the HFBTHO build system does not find \texttt{hdf5.mod} using \texttt{pkg-config}, locate \texttt{hdf5.mod} and add 
\texttt{-I/location/of/hdf5.mod} to the \texttt{OPTIONS\_FC} flag \hfbthoad's Makefile.

\subsection{Building and Running \hfbthoad}
Running a derivative simulation with \hfbtho using Tapenade is not very different from running any other typical \hfbtho simulation and broadly involves the same steps.

\begin{itemize}
    \item Set up environment variables

\begin{lstlisting}[language = bash, rulecolor=\color{white}]
$ export TAPENADE_HOME=/Location/of/Tapenade
$ export TAPENADE_LAPACK_SRC=/Location/of/LAPACK
$ export TAPENADE_LAPACK_INSTALLATION=\
         ${TAPENADE_LAPACK_SRC}/SRC/
\end{lstlisting}
    \item Go to the \verb|src| directory of HFBHTO. From the root of the HFBHTO repository, the path is as follows.

\begin{lstlisting}[language = bash, rulecolor=\color{white}]
$ cd src
\end{lstlisting}

    \item Edit \code{Makefile} to set variables.
\begin{lstlisting}[language = bash, rulecolor=\color{white}]
COMPILER    = IFORT
FORTRAN_MPI = ifort
USE_OPENMP      = 1
AD               = 1
AD_VECTOR_MODE   = 1
\end{lstlisting}

    \item Clean any remnant files in your build subdirectory from a previous simulation.

\begin{lstlisting}[language = bash, rulecolor=\color{white}]
$ make clean
\end{lstlisting}

\item Run the make commands.  The results in Section \ref{sec:results} are derived by using the forward mode with independent and dependent variables that are specified in the file \code{src/hfbtho_ad/MakefileTap}.

\begin{lstlisting}[language = bash, rulecolor=\color{white}]
$ make 
\end{lstlisting}

These commands will generate the executable \\ \code{hfbtho_ad/hftbho_main}.

\item Run the executable, and examine the output. 

If OpenMP is used, do the following first.
\begin{lstlisting}[language = bash, rulecolor=\color{white}]
$ export OMP_STACKSIZE="10 G"; export OMP_NUM_THREADS=32;
\end{lstlisting}
Then run the code using the provided input files.
\begin{lstlisting}[language = bash, rulecolor=\color{white}]
$ cd hfbtho_ad; 
$ ./hfbtho_main  -f hfbtho_UNEDF0_Z20_N20_sphGS.dat \
  -u unedf_UNEDF0_Z20_N20_sphGS.dat  > hfbtho_out.txt 2>&1
\end{lstlisting}

    \item Analyze the output written to \code{hfbtho_out.txt} as well as final $\rvec$ values written in \code{eresu.csv} and \code{ala2.csv}. The derivatives are written to \\ \code{eresu_d.csv} and \code{ala2_d.csv}.
\end{itemize}

\bibliographystyle{elsarticle-num-names}
\providecommand{\noopsort}[1]{}

\onecolumn
\framebox{\parbox{\columnwidth}{The submitted manuscript has been created by UChicago Argonne, LLC, Operator of Argonne National Laboratory (`Argonne'). Argonne, a U.S.\ Department of Energy Office of Science laboratory, is operated under Contract No. DE-AC02-06CH11357. The U.S.\ Government retains for itself, and others acting on its behalf, a paid-up nonexclusive, irrevocable worldwide license in said article to reproduce, prepare derivative works, distribute copies to the public, and perform publicly and display publicly, by or on behalf of the Government. The Department of Energy will provide public access to these results of federally sponsored research in accordance with the DOE Public Access Plan. \url{http://energy.gov/downloads/doe-public-access-plan}.}}

\end{document}